%% file: main.tex
\pdfoutput=1
\documentclass[journal]{vgtc}

\usepackage[table,xcdraw,dvipsnames]{xcolor}

\onlineid{0}

\vgtccategory{algorithm/technique}

\newcommand{\modelshort}{PAM}

\newcommand{\frameworkshort}{VADIS}

\newcommand{\revised}{\color{black}}

\vgtcpapertype{please specify}

\title{\emph{\frameworkshort{}}: A Visual Analytics Pipeline for Dynamic Document Representation and Information-Seeking}

\author{%
  \authororcid{Rui Qiu}{0000-0002-3905-8926},
  \authororcid{Yamei Tu}{0000-0002-0722-837X}, Po-Yin Yen and \authororcid{Han-Wei Shen}{0000-0002-1211-2320}
}

\authorfooter{
    \vspace{-4mm}
    \item Rui Qiu, Yamei Tu and Han-Wei Shen are with Ohio Sate University, 
E-mail: \{qiu.580, tu.253, shen.94\}@osu.edu; Po-Yin Yen is with Washington University School of Medicine, E-mail: yenp@wustl.edu.
  
}

\abstract{ In the biomedical domain, visualizing the document embeddings of an extensive corpus has been widely used in information-seeking tasks. However, three key challenges with existing visualizations make it difficult for clinicians to find information efficiently. First, the document embeddings used in these visualizations are generated statically by pretrained language models, which cannot adapt to the user's evolving interest.  Second, existing document visualization techniques cannot effectively display how the documents are relevant to users’ interest, making it difficult for users to identify the most pertinent information. Third, existing embedding generation and visualization processes suffer from a lack of interpretability, making it difficult to understand, trust and use the result for decision-making. 
In this paper, we present a novel visual analytics pipeline for user-driven document representation and iterative information seeking (VADIS). VADIS introduces a prompt-based attention model (PAM) that generates dynamic document embedding and document relevance adjusted to the user's query. 
To effectively visualize these two pieces of information, we design a new document map that leverages a circular grid layout to display documents based on both their relevance to the query and the semantic similarity. Additionally, to improve the interpretability, we introduce a corpus-level attention visualization method to improve the user's understanding of the model focus and to enable the users to identify potential oversight. This visualization, in turn, empowers users to refine, update and introduce new queries, thereby facilitating a dynamic and iterative information-seeking experience.
We evaluated VADIS quantitatively and qualitatively on a real-world dataset of biomedical research papers to demonstrate its effectiveness.

}


\teaser{
  \centering
  \includegraphics[width=\linewidth]{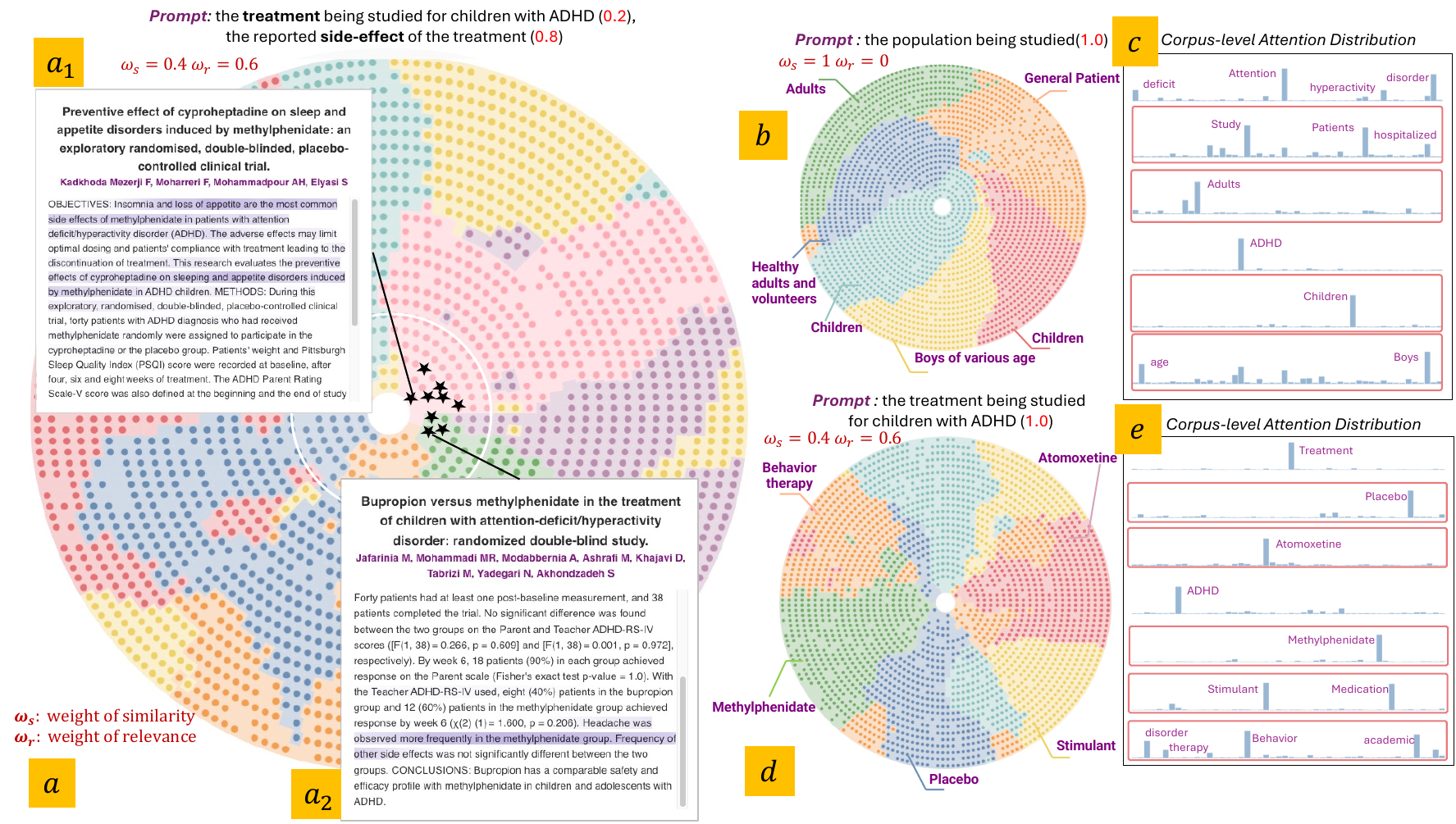}
  \vspace{-0.25in}
  \vspace{-0.06in}
  \caption{{\revised{Illustration of an information-seeking process where domain experts leverage VADIS to first understand the document distribution based on ``population being studied'' ($b$, $c$), and then shift to ``treatment being applied'' ($d$, $e$), and finally retrieve information with combined queries ``treatment being studied'' and ``the reported side-effect of the treatment'' ($a$, $a_1$, $a_2$). $w_s$ and $w_r$ are two parameters to customize the balance between relevance and similarity considerations when generating each circular document map.}}}
  \label{fig:teaser}
}

\graphicspath{{figs/}{figures/}{pictures/}{images/}{./}} 

\usepackage{tabu}                      
\usepackage{booktabs}                  
\usepackage{lipsum}                    
\usepackage{mwe}                       

\usepackage{mathptmx}                  

\usepackage{cite}                      
\usepackage{tabu}                      
\usepackage{booktabs}                  
\usepackage[ruled,vlined]{algorithm2e}
\usepackage{amssymb}
\usepackage{fontawesome}
\usepackage{dirtytalk}
\usepackage{tikz}
\usepackage{tabu}  
\usepackage{tabularx}
\usetikzlibrary{shapes}
\usepackage{makecell}
\usepackage{multicol}
\usepackage{multirow}
\usepackage{colortbl}
\usepackage[most]{tcolorbox}
\usepackage[framemethod=TikZ]{mdframed}
\usepackage{lipsum}
\usepackage{stfloats}
\usepackage{titlesec}
\usetikzlibrary{plotmarks}
\newcommand{\tikzcircle}[2][red,fill=red]{\tikz[baseline=-0.5ex]\fill[#1,radius=#2] (0,0) circle ;}%
\usepackage{enumitem}

\begin{document}



\maketitle

\input{tex/1Introduction}
\input{tex/2RelatedWork}
\input{tex/3Background}
\input{tex/4Requirement}
\input{tex/5Approach-PAM}
\input{tex/6CaseStudy}
\input{tex/7_discussion}
\bibliographystyle{abbrv-doi}
\bibliography{template}
\end{document}

%% file: tex/1Introduction.tex
\firstsection{Introduction}
Seeking information from large collections of documents is a critical challenge across many domains. It is particularly pressing in the biomedical field, where new research is published at an unprecedented rate, and clinical decisions rely on up-to-date research evidence. Recent advancements in transformer-based language models~\cite{transformer}
have shown significant promise in aiding information-seeking through various tasks, such as encoding rich semantic information into embeddings and generating comprehensive summarization. 

\begin{figure}
    \centering
    \includegraphics[width=\linewidth]{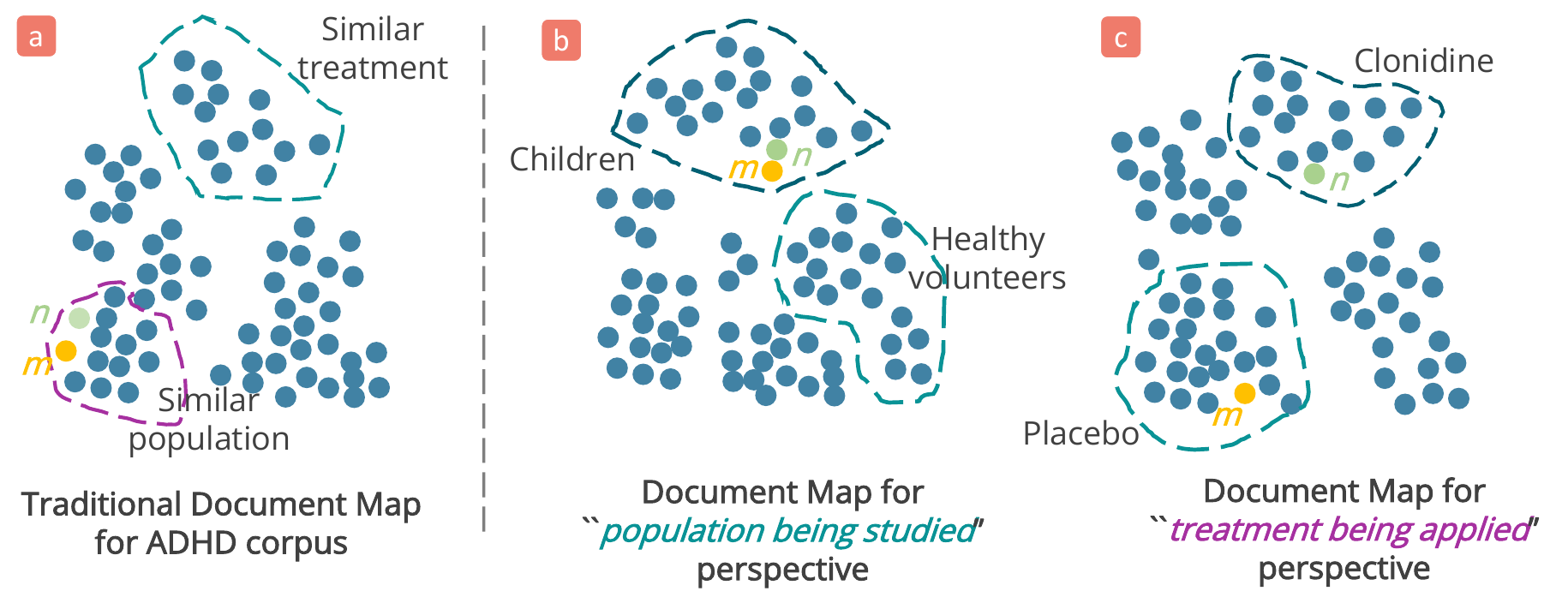}
    \vspace{-5mm}
    \caption{Illustration of traditional document map compared with document map for a specific perspective. $m$ and $n$ are two documents that study the same population, as shown in figure $b$, but with different treatments being tested, as shown in figure $c$.}
    \label{fig:intro-embedding}
    \vspace{-5mm}
\end{figure}

Despite these advancements, the specialized nature of biomedical research requires intensive human involvement to ensure accuracy and trustworthiness \cite{khullar2024large}. 
Given this context, our research is centered on supporting clinicians in iterative exploring and retrieving documents of their interest using transformer-based language models while keeping them aware of the model's decision. We start by identifying four major challenges that existing works failed to address. \emph{Firstly}, existing methods often generate static document representations (e.g., embeddings) that do not align with the user's interests. The visualization produced using such representations, along with commonly used scatterplot-based document maps, will exhibit poor distribution and can confuse users attempting to navigate and explore the data. For instance, when searching for ADHD (attention deﬁcit hyperactivity disorder) studies via visualizing the distribution of similar ADHD studies, 
with static embedding approaches, some documents (studies) may be grouped together because of their similar study populations (Figure \ref{fig:intro-embedding}-a), while others may be clustered due to similar drug treatments. This can hinder users from identifying specific research studies if their primary interest is in treatments only. \emph{Secondly}, when a user submits a query, it's likely that not all documents within the corpus will align with their specific interests. However, existing document visualizations, e.g., document maps projecting document embeddings, only focus on showcasing the distribution of documents based on semantic similarity, and do not present documents relevant to user's query. This limits clinicians' ability to efficiently navigate and identify relevant documents from the corpus. \emph{Thirdly}, interpretability of resulting embeddings poses a significant issue. As the size of transformer models increases, visualizing all attention layers and heads to understand their distribution become impractical, making it difficult to assess the embedding representation. \emph{Fourthly}, the iterative nature of the information-seeking process is often under-supported by current visualizations and visual analytics systems. In real-world scenarios, clinicians often ask a series of questions and engage in an iterative searching process. They are likely to generate new interests or adjust based on the searched findings. However, existing methods offer limited support for this iterative exploration, failing to update document representations to reflect users' evolving interests. 

To address these challenges and tackle the complexity inherent to information-seeking, in this paper, we propose a \textbf{V}isual \textbf{A}nalytics Pipeline for Dynamic \textbf{D}ocument Representation and \textbf{I}nformation \textbf{S}eeking (VADIS). VADIS consists of three integrated components. 
\emph{Firstly}, considering the limitation of existing document maps, we propose a Prompt-based Attention Model (PAM) that generates dynamic document representations aligned with user's evolving prompts or interests, and computes relevance scores to these prompts. The PAM model serves as the foundation for VADIS to construct a novel relevance-preserving document map. The map adopts a circular layout to project documents with higher relevance closer to the center, symbolizing user's interest, while the spatial distance between documents reflects their semantic similarity. This arrangement allows users to easily grasp both the semantic landscape of documents and their pertinence to the search queries. 
\emph{Secondly}, the design of VADIS prioritizes interpretability to enhance the user's understanding of PAM's attention to queries. VADIS calculates and visualizes the model's attention focus at the corpus level, not only at the individual document level. This visualization alleviates the user's burden of analyzing attention across individual documents to infer attention's focus areas, thus providing clearer and more interpretable search results in response to user queries. 
\emph{Thirdly}, a fundamental aspect of VADIS is its support for the iterative nature of information-seeking. VADIS enhances users' understanding of the corpus and model's current focus via a document map and attention visualization, thereby facilitating the formulation of new queries or the refinement of existing queries. Upon receiving new or modified queries, VADIS recalculates the document representation and relevance scores with PAM, subsequently updating the relevance-preserving document map and attention visualizations. This support users's continuous engagement in a dynamic exploration of the document corpus. Overall, the main contributions of our work are as follows:

 \setlist{nolistsep}
\begin{itemize}
    \item We propose a visual analytics pipeline (\frameworkshort{}) for dynamic document representation and information seeking.
    \item We propose Prompt-based Attention Model (PAM) for generating dynamic document representation and calculating relevance score based on user-driven interests. 
    \item We propose a Relevance-preserving Document Map for visual encoding of document relevance and semantic similarity, along with a corpus-level attention visualization for attention interpretability. 
    \item We demonstrate the effectiveness of \frameworkshort{} through quantitative evaluation and case studies.
    
\end{itemize}

%% file: tex/2RelatedWork.tex
\section{Related Work}
\subsection{Visual Analytics System for Information-Seeking}
Visual analytics systems for information-seeking typically leverage text-mining approaches and interactive visualizations, demonstrating their effectiveness in tasks such as document retrieval. Existing systems can be divided into two categories: the first focuses on generating static overviews of document collections, while the other strives to improve the entire information-seeking workflow. In the first category, related works VOSViewer\cite{vosviewer} supports information seeking on literature search by providing overviews of reference and citation relationships. CiteRiver\cite{citeriver} links topic and venue citations, while Bridger\cite{bridge} generates an overview of authorship by performing semantic authorship representations. Other studies \cite{keyvis, scitool, evidenceSet, architext} 
focus on keywords and topic perspectives. More recently, with the development of NLP, research studies have started to focus on visualizing document embedding to provide a corpus overview. Related works TRIVIA\cite{dias2019trivir} and Vitality\cite{narechania2021vitality} leverage advanced language models to generate text embeddings and project into 2D scatter plots to display the distribution of corpus, which help users find relevant research articles (papers) by identifying similarities with the embedding space. In contrast to these existing works that produce static overviews, VADIS stands out by offering dynamic overviews tailored to users' interests. This unique approach ensures that users receive customized visualizations that cater to their specific needs, resulting in a more effective overview. In the second category, systems 
\cite{ponsard2016paperquest,visualbib,allot2019litsense,choe2021papers101}
streamline the systematic review process with interactive knowledge extraction, recognition, and retrieval. These tools offer various views and features for visualizing and adapting results while also supporting document discovery and organization through multiple visualization panels. However, these existing systems often fall short in offering users insight into the underlying rationale of these outcomes. This limitation can restrict users' ability to refine or expand their information-seeking activities. Recent developments, like DocFlow, have attempted to grant users more control by allowing the construction of custom retrieval pipelines with different perspectives. While this adds a layer of flexibility, it still lacks in areas such as presenting diverse perspectives and help domain experts understand the model's decision. VADIS distinguishes itself by offering not just an overview but also a clear interpretability of the model's attention mechanism, facilitating a more intuitive and iterative information seeking pipeline.
\vspace{-2mm}
\subsection{Document Representation and Retrieval}
Document representation is a classic problem in natural language processing (NLP) and information-seeking \cite{teller2000speech}. It aims to convert textual data into vector representation that captures the meaning. Two main approaches have been proposed for generating document embedding. The first approach focuses on the lexical properties of the text, creating representations based on term frequencies, such as Bag-of-Words \cite{zhang2010understanding}, TF-IDF\cite{tfidf}, and BM25\cite{bm25}. The second approach captures the semantic meaning of the text using advanced techniques, including deep learning models like Word2vec \cite{word2vec}, Doc2vec \cite{doc2vec}, and GloVe \cite{glove}. More recently, attention-based language models, such as BERT\cite{bert}, ALBERT\cite{roberta}, and GPT\cite{GPT-3}, have demonstrated superior performance across a wide range of NLP tasks. These models leverage the transformer architecture and self-attention mechanisms to generate semantically rich document embeddings. The quality of document embedding directly impacts the effectiveness of document retrieval. Existing methods leverage embeddings to retrieve documents by calculating document similarity or measuring the similarity between documents and user queries. Examples of these techniques include DSSM \cite{DSSM}, CLSM\cite{CLSM}, DESM\cite{DESM}, DPR\cite{DPR}, and QDR\cite{Docflow}. These methods encode queries and documents, then compute their similarity to determine relevance. Among these approaches, DPR and QDR employ a dual encoder structure to encode queries and documents separately, resulting in higher retrieval accuracy. Compared with these works, which generate static document embeddings solely for retrieval purposes, we leverage token embeddings from the documents. Our approach offers two main benefits: first, incorporating token information provides more detail and leads to higher accuracy; second, using token information and prompts allows us to compute prompt-based document embeddings, offering a more customizable representation tailored to the user's interests.

%% file: tex/3Background.tex
{\revised
\vspace{-1mm}
\section{Background} \label{SEC:BACKGROUND}
\subsection{Attention Mechanism in Transformer}
The transformer model \cite{transformer} is a neural network architecture that has revolutionized the field of Natural Language Processing by providing an effective way to process sequential data. 
A key component of the transformer model is the attention mechanism, which allows the model to focus on specific parts of the input sequence during processing. 
Considering BERT\cite{bert} as an example, it is a multi-layer transformer encoder. Each layer has multiple attention heads that compute the contextualized embedding for each token by incorporating information from other tokens. 
A special [CLS] token is added at the beginning of each input to represent the entire sequence. 
Generating document embedding is performed by computing the contextualized embedding for the  [CLS] token and using its embedding to represent the entire input sequence. 
Specifically, the attention mechanism calculates an attention score between each token in the input sequence and the $\text{[CLS]}$ token using query ($W^Q$), key ($W^K$), and value ($W^V$) matrices, which are learned during the training process. The attention score from the [CLS] token to other tokens in the sequence is given by
\begin{equation}
    attn(\text{[CLS]}, :) = \text{softmax}\left(\frac{e_{\text{[CLS]}}^\top {W^Q} {W^K}^\top E} {\sqrt{d_k}}\right),
\end{equation}
where $d_k$ is the dimensionality of the query and key vector, and $e_{\text{[CLS]}}$ and $E$ are the embeddings for the $\text{[CLS]}$ token and the embedding matrix of the input tokens, respectively.
The attention scores are then used to compute a weighted sum of the value embeddings, which gives the embedding for the $\text{[CLS]}$ token:
\begin{equation}
    e_{\text{[CLS]}} = attn(\text{[CLS]},:)^\top ((W^V)^\top E)
\end{equation}
In this paper, prompt-based attention model (PAM) leverages this attention mechanism to dynamically assign higher attention scores to tokens relevant to the prompt, thus focusing more precisely on the pertinent information. This process is analogous to how BERT uses the [CLS] token to capture global semantics, but PAM is specifically trained to enhance the relevance of embeddings based on user prompts.
}



{\revised
\subsection{Prompt-tuning}
In the domain of language models, prompt-tuning, also known as prefix-tuning, refers to a technique used in training to generate relevant text to a specific prompt. This is achieved by adding a special token at the beginning of the input sequence to indicate the desired behavior. By fine-tuning the language model on the specific prompt, it becomes more targeted in its output, leading to more accurate and coherent text generation in particular domains or contexts. Prompt-tuning has been successfully implemented in many popular language models for multi-task learning. For instance, in T5\cite{t5} and GPT-3\cite{GPT-3}. 

We leverage the concept of prompt-tuning in the design of the Prompt-based Attention Model (PAM). Instead of generating relevant text, we fine-tune the model’s attention mechanism to focus more on information relevant to the prompt. 
Specifically, PAM uses the query as the special token to guide model's attention and fine-tunes the query ($W^Q$), key ($W^K$), and value ($W^V$) matrices to ensure that the attention scores highlight the tokens most pertinent to the prompt, thereby refining the contextual embeddings generated by the model.

}

%% file: tex/4Requirement.tex
\section{Requirement Analysis} 
We conducted a preliminary study with two domain experts ($E1$ and $E2$) with related experience to ensure the validity of our analysis. Specifically, $E1$ is a clinical expert with extensive experience in information retrieval, information seeking, and systematic review in the biomedical informatics domain. $E2$ is an expert in information visualization and retrieval with a computer science background. We held regular meetings with the participants to evaluate the strengths and limitations of existing information-seeking systems and visualizations. The discussions also delved into the ways in which each visualization is generated and the methods used. Based on the feedback received, we identified four primary requirements for our proposed method and system design:

\emph{R1: Provide an overview that reflects document distribution based on user interest.} $E1$ stated that in many systematic review tasks, researchers need to explore studies based on different perspectives, such as treatment or population being studied in biomedical research.
Therefore, the proposed visualization should enable users to easily identify relevant clusters based on their interests and explore the relationships between different clusters. 
$E2$ further added that a successful map-based visualization depends on a meaningful similarity measure. However, two documents can be similar from one perspective, like the population being studied (Figure \ref{fig:intro-embedding}-$b$), but dissimilar from another, like treatment being applied (Figure \ref{fig:intro-embedding}-$c$). Therefore, developing a user-driven document embedding method that can capture various perspectives and generate customized document maps would greatly improve the usability and effectiveness of map-based visualization for information-seeking. 

\emph{R2: Highlight relevance on the map and enhance the clarity of visualization}. Besides the user-driven distribution, E1 emphasized the need for visualizations to highlight the relevance of documents to the users' interest, particularly in the context that only a fraction of the articles are pertinent. 
Moreover, E1 highlighted the issue of visual clutter in traditional scatterplot-based document maps, which obscure the visibility of individual documents and prevent clinicians from efficiently identifying relevant information.
These concerns call for a visualization that projects documents based on their relevance and similarity while displaying them in a clearer and organized manner.

\emph{R3: Ensure model interpretability for domain experts.} Domain experts expressed the need to understand the information processed by the model in order to make informed decisions. This requirement highlights the significance of ensuring the model's insights are interpretable, thereby enhancing trust and applicability in clinical settings. 

\emph{R4: Support iterative exploration.} Our discussions with domain experts revealed that information-seeking is often an iterative and complex process. E2 pointed out that it is critical to support an iterative exploration, as researchers need to explore different perspectives and dive deeper into the data as their understanding evolves. E1 shared similar views that in the biomedical domain, clinicians often need to identify and categorize treatments for diseases and related populations. As they explore and organize the findings of these perspectives, more nuanced inquiries often emerge, driving clinicians to delve deeper into specific aspects of their research. This iterative process of refining questions and seeking out new relevant studies necessitates a flexible pipeline that not only facilitates the exploration of diverse perspectives but also promotes the identification of the most relevant studies.

To fulfill the proposed requirements, we recognize the need to combine efficient document representation methods with an intuitive visualization for an optimal solution. In the following sections, we will first present our approach to provide a user-driven document representation for \emph{R1}, followed by the design and development of our visualizations and mapping algorithms for \emph{R2}, \emph{R3} and \emph{R4}.


%% file: tex/5Approach-PAM.tex
\section{VADIS for iterative information-seeking}
This section introduces the components of our proposed framework VADIS. We start by outlining the Prompt-Based Attention Model (PAM) in \autoref{sec:pam}, which is the core of our framework.
The section details PAM's architecture and how it manages to generate query-specific document representations and document relevance. In \autoref{sec:map}, we introduce the relevance-preserving document map, which integrates with PAM to visually represent documents considering their relevance and semantic similarity. \autoref{sec:attentionVis} delves into the visualization of corpus-level attention. It aims to enhance PAM's interpretability, allowing users to understand the model’s focus so as to identify possible overlooked areas. Lastly, \autoref{sec:iterative} demonstrates the system's iterative adaptability, showcasing how user feedback refines PAM-generated data and subsequent visualizations.
\begin{figure}
    \centering
    \includegraphics[width=1.0\linewidth]{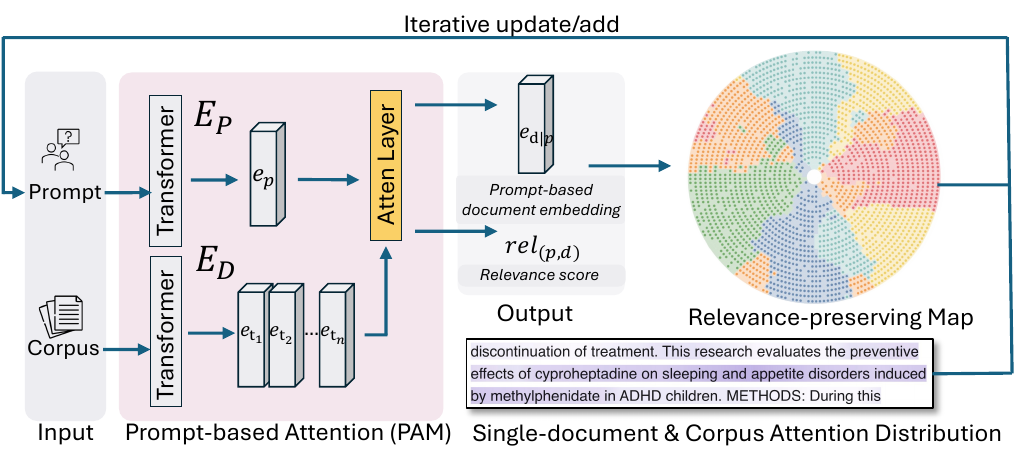}
    \caption{The illustration of VADIS's iterative pipeline, which consists of the prompt-attention model and subsequent visualizations.}
    \vspace{-3mm}
    \label{fig:pipeline-overview}
\end{figure}
\vspace{-1mm}
\subsection{Prompt-Based Attention Model} \label{sec:pam}
In this section, we discuss how the Prompt-based Attention Model generates dynamic embeddings according to user-provided prompts. 
This is accomplished through a weighted sum of token embeddings, where the weights are formalized as attention scores calculated from each token to the user's prompt. We will introduce the designed structure of PAM and how it is trained to achieve the desired goal.

\subsubsection{Text Representations with Encoders}
Given the fact that \emph{prompts} and \emph{documents} may differ in length, tone, and semantics, we employ two separate encoders, denoted as $E_P(\cdot)$ and $E_D(\cdot)$ and train them jointly to capture their distinct characteristics. 
To increase the adaptability of the PAM in various scenarios, $E_P(\cdot)$ can be any language model that generates a sentence embedding for the prompt, denoted as $\textbf{e}_p$, while $E_D(\cdot)$ can be an arbitrary pretrained transformer that generates the embedding for each token $i$ inside each document $j$, denoted as $\textbf{e}_{t_{ij}}$. The token embedding already captures contextual information from its surrounding tokens in the transformer architecture. Also, $\textbf{e}_p$ and 
$\textbf{e}_{t_{ij}}$ should share the same dimensions $\textit{dim}$.

\subsubsection{Attention Layer}
To learn the proper relationships between each prompt and relatively important tokens, we employ an additional attention
layer built on top of the two encoders (Fig.\ref{fig:pipeline-overview}). 
While self-attention has been widely used in transformers, users often find it hard to interpret the patterns from multiple heads across multiple layers. 
In contrast, the adoption of our attention layer explicitly learns the overall relationships between prompts and tokens, which has several follow-up usages: 
(1) the attention pattern is distilled and concise, which provides a way for interpreting the embedding generation. 
(2) the token-level attention can be aggregated to measure the relatedness of each document, assisting in filtering irrelevant documents. 
(3) the attention can be utilized to weight token embeddings to generate dynamic document embeddings. 

The attention layer has two learnable matrices $\textbf{W}^Q$ and $\textbf{W}^K$. They are utilized to convert prompt and token embeddings to $query$ and $key$ vectors, which later can perform the dot product to measure the relationship strength as attention. The attention between prompt $p$ and token $j$ in document $i$ can be formulated as
\vspace{-1mm}
\begin{equation}    
    attn(p, t_{ij}) = (\textbf{W}^Q \textbf{e}_p )\cdot (\textbf{W}^K\textbf{e}_{t_{ij}}).
    \vspace{-1mm}
\end{equation}
During training, the model learns to assign higher scores to more relevant tokens, allowing the model to generate a more accurate embedding. 

Retrieving relevant documents based on prompts is an important task in information seeking, typically accomplished by computing the similarity between the embeddings of the prompt and documents. This process can be effectively conducted using the attention scores, which significantly simplifies the computational complexity. To achieve this, we define the relevance of a document $d_j$ with respect to a  prompt $p$ as the sum of attention scores across  all tokens within $d_j$ in relation to $p$, formulated as follows:
\vspace{-2mm}
\begin{equation}
    r(p, d_j) = \sum_{t_{ij} \in d_j}attn(p, t_{ij}).
    \vspace{-2mm}
\end{equation}

Furthermore, the attention scores facilitate the computation of dynamic document embeddings through a weighted sum approach. Specifically, we first apply the softmax function to normalize these scores, transforming them into a set of weights. Then, these weights are applied to the respective token embeddings, enabling the creation of a document embedding that is contextually aligned with the prompt. The calculation can be mathematically expressed as
\vspace{-1mm}
\begin{equation}
    \textbf{e}_{d|p} = \sum_{t_{ij}\in d_j}softmax(\frac{attn(p, t_{ij})}{\sqrt{dim}})\textbf{e}_{t_{ij}}.
    \vspace{-1mm}
\end{equation}



\subsubsection{Loss Functions}
The training objective of the PAM is to enhance models' ability to accurately attend to tokens based on their relevance to a given prompt. 
However, instead of directly training the attention between the prompt and tokens, which is intrinsically challenging, we train the relevance between the prompt and documents to indirectly learn the token-level attention scores. The reason is that document-level relevance is computed as a sum of attention scores between the prompt and tokens; thus, training the relevance implicitly optimizes the attention scores. 

To accomplish this goal, we employ contrastive learning \cite{infoNCE} to force the relevance score to capture the true relationships between documents and the prompt. Contrastive learning operates by comparing a ``positive'' example (a document relevant to the prompt) against a set of ``negative'' examples (documents irrelevant to the prompt). Specifically, given a prompt $p$, we prepare a collection of documents $D = \{d^+, d_1^-, d_2^-, ..., d_{n-1}^-\}$,  where $d^+$ is a relevant document, and $d_i^-$ represents an irrelevant document. These irrelevant documents are randomly sampled from the document pool, excluding the relevant ones. 
The training process is formulated to minimize the following loss:
\vspace{-1mm}
\begin{equation}
    Loss(p, d^+, d_1^-, ... d_{n-1}^-) = -log \frac{e^{r(p, d^+)}}{e^{r(p, d^+)} + \sum_{i=1}^{n-1}e^{r(p, d_i^-)}}\text{}.
    \vspace{-1mm}
\end{equation} 
During training, the attention between the prompt and positive documents $attn(p,d^+)$ will be increased, and attention to the negative ones $attn(p,d_i^-)$ will be decreased. 

\subsubsection{Implementation Details}
We trained the PAM model in the biomedical domain, which involved several important considerations, including  
the selection of model architecture, training datasets, and hyperparameter settings for the PAM model.
\begin{itemize}
    \item \emph{Model Architecture}: we employ BioBERT \cite{biobert} to initialize both $E_p(\cdot)$ and $E_d(\cdot)$. 
BioBERT was pre-trained on a large corpus of 
biomedical literature, ensuring its understanding of domain-specific language and terminology. 
\item \emph{Dataset Selections}: To ensure that the model learns both general and domain-specific knowledge, we leverage four datasets:   SQuAD\cite{squad}, TriviaQA\cite{triviaqa}, NaturalQA\cite{wikiqa}, and emrQA\cite{emrqa}. In particular, emrQA is a medical QA dataset consisting of over 27,000 question-answer pairs sourced from clinical notes and discharge summaries, making it particularly relevant for information-seeking in the biomedical domain. 
\item \emph{Dataset Transformation}: To prepare the training data for contrastive learning, we transform several existing extractive question-answering (QA) datasets. These datasets consist of (\emph{question}, \emph{answer}, \emph{context}) triplets. We use the \emph{question} as the prompt $p$, the corresponding \emph{context} as the positive documents $d^+$, and randomly sample $n-1$ contexts from other triplets as negative documents $\{d_1^-, d_2^-, ..., d_{n-1}^-\}$. The   training data is formulated as ($p$, $d^+$, $\{d_1^-, d_2^-, ... d_{n-1}^-\}$) with $n=16$. 
\end{itemize}

\vspace{-2mm}
\subsection{Relevance-preserving Document Map}\label{sec:map}
Building on the embedding and relevance scores modeled by PAM, our relevance-preserving map is designed to project documents into a 2D space based on these two features (Fig.\ref{fig:pipeline-overview}).
In this section, we elucidate the design logic and the corresponding mapping algorithm of the relevance-preserving document map.

\subsubsection{Design Logic}
\begin{figure}
    \centering
    \includegraphics[width=.9\linewidth]{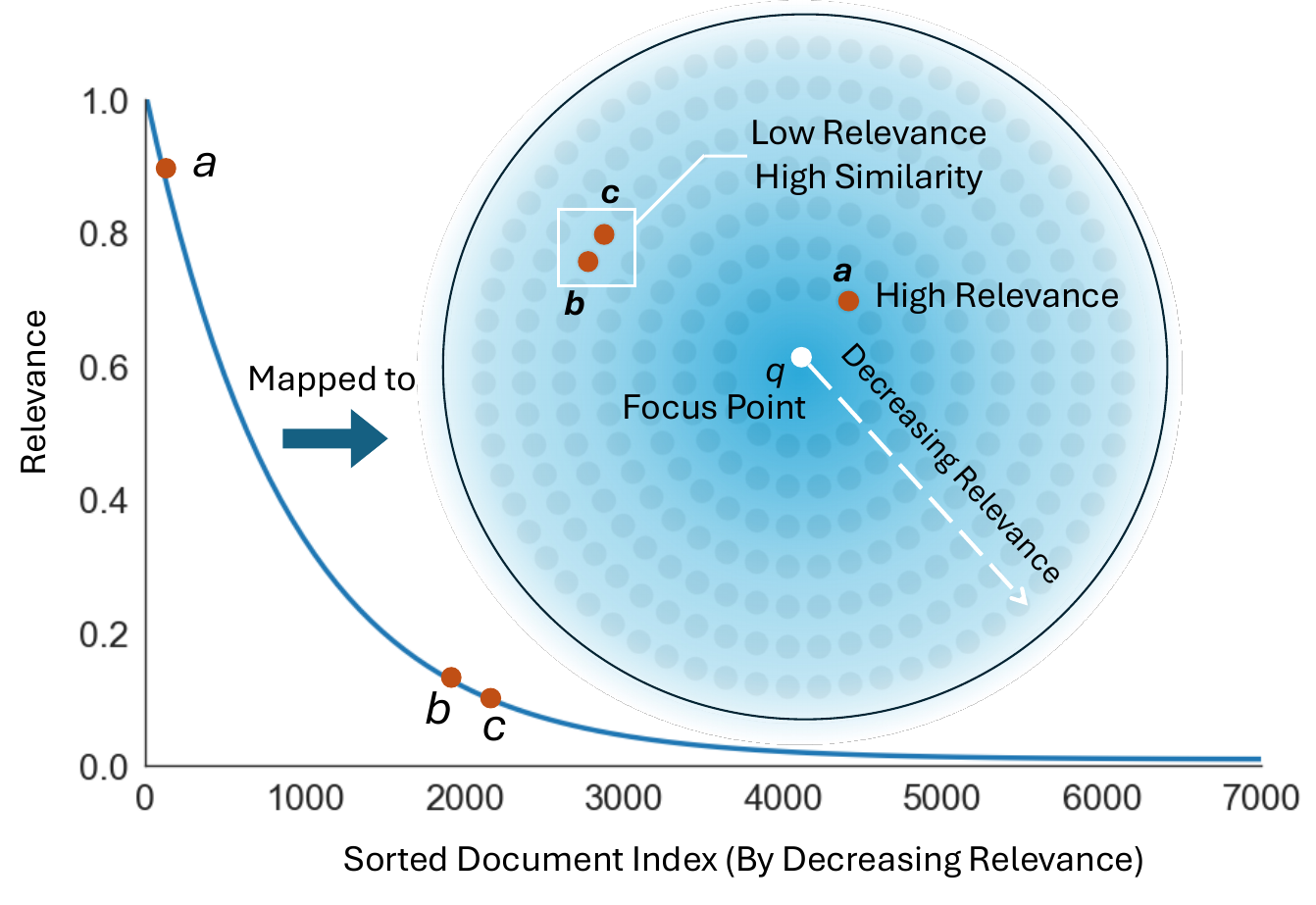}
    \caption{Relevance of documents in a corpus based on query ``\emph{side-effect of methylphenidate for Children with ADHD.}''}
    \vspace{-4mm}
    \label{fig:motivation}
\end{figure}

The design of the relevance-preserving document map integrates a circular grid layout (Fig.\ref{fig:motivation}) which projects documents with a dual consideration of relevance and similarity. 
This design choice stems from three integrated considerations. \emph{First}, empirical studies in the biomedical domain (such as the one illustrated in Fig.\ref{fig:motivation}) demonstrate that within a vast corpus, only a small subset of documents holds substantial relevance to a given query. This pattern circle is inherently smaller, it highlights the limited number of highly relevant documents, whereas the progressively expanding peripheral zones are capable of accommodating the increasing volume of less relevant ones. Therefore, we adopt the circular layout where the center of the map is conceptualized as the query's focus point, and the distance between documents to the center reflects its relevance (Fig.\ref{fig:motivation}). \emph{Second}, the map should inherently display the semantic relationship between documents. This requires the position of documents not only consider the proximity to the center but also the relative distance to each other. \emph{Third}, to overcome the problem of over-clutter in traditional document maps, we adopt a grid layout inside the circular to increase the clarity of the projection, which has proved to be superior in displaying intensive information \cite{phrasemap}.  


\subsubsection{Relevance-preserving mapping}
To effectively map documents into our novel circular grid layout, we tackle the challenges of finding the best cell for each document, such that it can minimize a global loss that integrates both the document relevance and semantic similarity. 
Specifically, we consider the mapping should ensure that once a document is assigned to a particular cell, its neighboring cells should contain documents with similar embeddings, thereby preserving a pattern of semantic similarity across adjacent spaces.
At the same time, the relevance of documents must align with their spatial positioning, ensuring that each layer's relevance matches the documents it contains. 
This dual-focus mapping on grid layout transforms the challenge into mapping from a continuous space to a discrete one, thus we cannot leverage gradient descent to find the optimal position for each data point.


To address the challenge, we consider the mapping necessitating a two-step process: initially, documents compete for positions within the grid based on their relevance and similarity scores; subsequently, the grid should undergo an update step to refine the placement, enhancing both semantic coherence and relevance alignment. 
Such a process closely aligns with the principle of the competitive learning mechanism in Self-Organizing Maps (SOM) \cite{som}, which projects high dimensional data in continuous space to discrete grid cells. Specifically, SOM assigns each grid cell with a weight vector $w_j$ that matches the data dimensionality. This weight vector serves as the representation of each cell. Throughout the training, the algorithm assigns the data to the best matching cell (BMC) solely based on the similarity between data and the cell's weight vector (competition process), subsequently updating the weight vectors of BMC's neighboring cells to more accurately reflect the underlying data distribution (cooperation process). 

To align with our objectives, we extend the SOM with three modifications: (1) the introduction of a relevance parameter $\gamma_j$ to represent each cell's relevance, which plays a critical role in guiding data assignment and also updates during the training to reflect a more accurate distribution of data relevance, (2) a one-to-one mapping strategy between
grid cells and documents to increase the clarity, diverging from the original SOM where a single cell may represent multiple items and (3) an updated loss function for the competition step that for each document, find the best-fit cell that gives minimum weighted distance with respect to the similarity and relevance:
\vspace{-1mm}
\begin{equation}\label{eq:loss}
    L = \sum_{d=1}^{N} \underset{j}{min} ~\omega_s \left \|e_d - w_j\right \|_2 + \omega_r\left \|r_d - \gamma_j\right \|_2,
    \vspace{-1mm}
\end{equation}
where $\omega_r$ and $\omega_s$ are the coefficients assigned for the relevance measure and similarity measure.
$e_d$ and $r_d$ are the embedding and relevance of document $d$, respectively.
The detailed process is as follows (Fig.\ref{fig:pipeline}): 
\begin{figure}
    \centering
    \includegraphics[width=\linewidth]{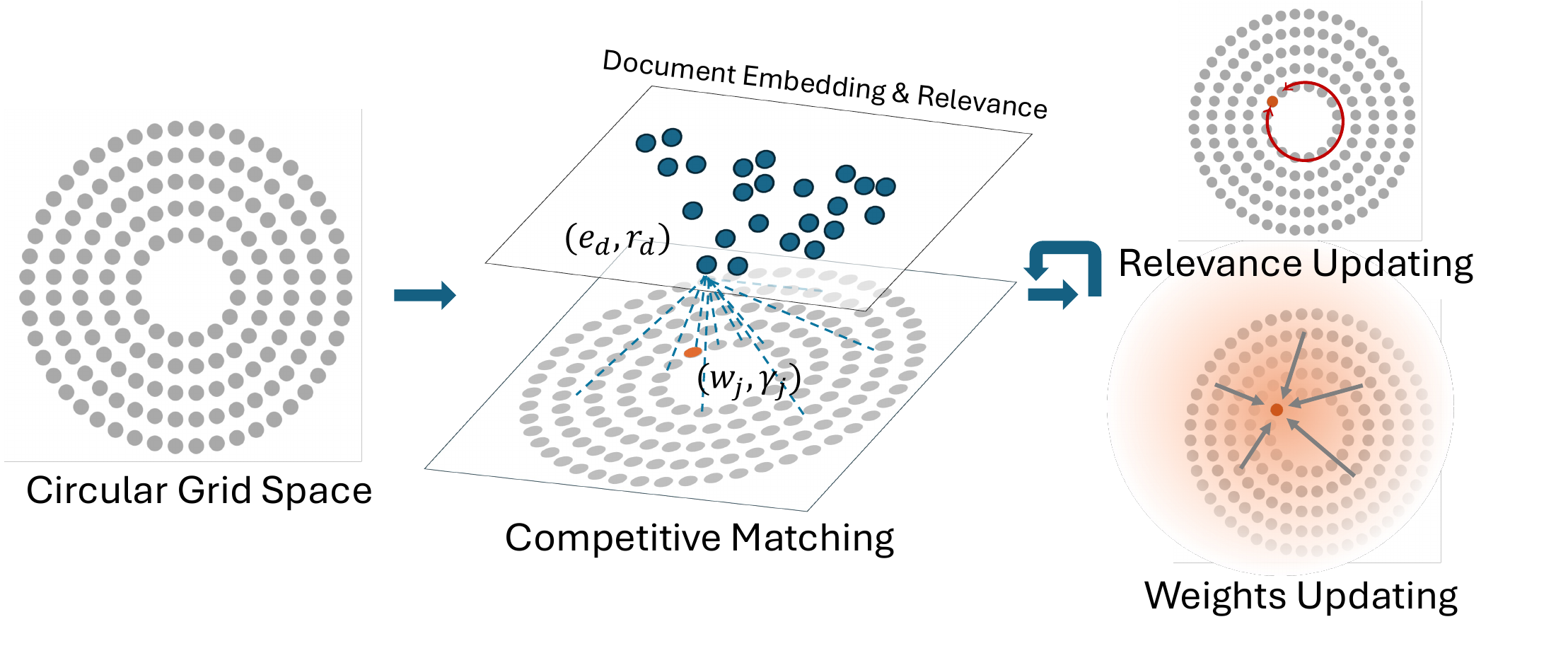}
    \vspace{-5mm}
    \caption{The learning pipeline of the relevance-preserving mapping}
    \label{fig:pipeline}
    \vspace{-5mm}
\end{figure}

\textit{Step 1. Initialization}. Given $n$ documents with embedding and relevance pairs $(e_1, r_1),...(e_n, r_n), e_i \in \mathbb{R}^{768}$, initialize $m$ grid cells ($m > n$) in a way that the innermost layer is populated with an initial number of points and each subsequent outward layer incrementally increase the number of cells. 
We deliberately leave the center as vacant to denote the query's focus point. Each cell is assigned with a random weight $w_j, w_j \in \mathbb{R}^{768}$ and a relevance $\gamma_j$, which is initialized to be inversely proportional to the distance between each cell and the center, and then normalized across all cells to the range [0, 1].

\textit{Step 2. Competitive Matching}. For each target data pair $(e_d, r_d)$, assign it to the vacant cell $c_j$ in where the \autoref{eq:loss} is minimized.

\textit{Step 3. Weights \& Relevance Updating}. To ensure that similar documents are mapped in close proximity, neighboring cells are expected to have similar weights $w$. We follow the approach in SOM that when a cell $j$ is assigned to a document, both the cell and its neighboring cells $k$ will be updated according to the topological distance using
\begin{equation}
    w_k = w_k + lr(t)\cdot T_{k, j}(t) \cdot (e_d - w_j), 
\end{equation}
where $lr(t)$ is the learning rate at epoch $t$ and $T_{k,j}(t)$ indicates the scale of the update based on the distance between $k$ and $j$, e.g, a Gaussian Kernel function centered at the cell $j$. This scale decreases as epoch $t$ increases. 
In the meanwhile, considering that the initial relevance assigned to each grid cell may not precisely align with the actual distribution of data relevance, we also update the relevance for all cells in the same grid layer of the cell $j$ with 
\begin{equation}
    \gamma_k = \gamma_k + lr(t)\cdot TR_{k, j} \cdot (r_d - \gamma_j), 
\end{equation}
where $TR_{k,j}$ indicates the scale of updates based on the topological distance between cell $k$ and $j$. Specifically, $TR_{k,j} = 1$ when $k$ and $j$ are in the same layer and $TR_{k,j} = 0$ otherwise. This ensures that each layer has a consistent relevance. By iteratively update the relevance for each layer, the collective relevance of the layer become more closely corresponds with the actual relevance value of the corpus. 

\textit{Step 4. Continuation}. Repeat steps 2-3 until all documents are assigned. Before the start of a new epoch, mark all cells as vacant.

By following these four steps, the mapping algorithm learns to assign documents to a structured circular grid by balancing the intricacies of relevance and semantic similarity.

\vspace{-1mm}
\subsection{Corpus-level Attention Visualization}\label{sec:attentionVis}
While the relevance-preserving map presents the distribution of documents based on PAM, it is necessary for users to understand how the embeddings are generated and especially, what information is being attended during the generation. However, analyzing the attention patterns across the entire corpus is cumbersome. As attention is calculated for each input article independently, users typically need to examine the attention distribution for every single article in order to get an overview. Such analysis is labor-intensive, especially when dealing with a corpus with a vast number of documents. 



To provide users an overview of the attention's focus, we introduce \emph{corpus-level attention}, which is the aggregated latent pattern of attention across the entire corpus. Upon analyzing the hidden patterns, we find that these aggregated attention pattern closely resembles the topics of a corpus that generated by topic modeling methods.
Specifically, we consider PAM's attention score is particularly suited to topic modeling based on two key observations. \emph{Firstly}, the attention mechanism matches the generative view of topic modeling. PAM exhibits a selective attention that can adjust its focus based on different prompts. 
This ability suggests that its attention mechanism can identify and align with prompt-specific interests or topics. Consequently, the attention scores observed across documents can be interpreted as how the prompt-driven interests selectively attend to the most relevant content that reflect its current focal point. 
Such a process closely resembles the generative view of topic modeling methods, where documents are conceived as mixtures of several topics, each defined by a distinct distribution of words. \emph{Secondly}, the topic modeling result of the attention can reflect more accurate query-based topics. Attention scores from PAM serves as a more precise measure of token importance. Compared to assessing the token importance by counting its frequency throughout the corpus, attention scores capture the contextual significance of each token in relation to the user's prompts or queries. 

Given the non-negativity of attention scores, we apply Non-negative Matrix Factorization (NMF) to extract the topics of attention. NMF is a technique widely used in topic modeling that decomposes a documents-token frequency matrix $\textbf{V}$ into two interpretable matrices by minimizing
\begin{equation}
    \Vert \textbf{V}-\textbf{WH} \Vert_F,
\end{equation} 
where $\textbf{W}$ captures distribution of latent topics across documents and $\textbf{H}$ details the contribution of each token to these topics.

\begin{figure}
    \centering
    \includegraphics[width=0.9\linewidth]{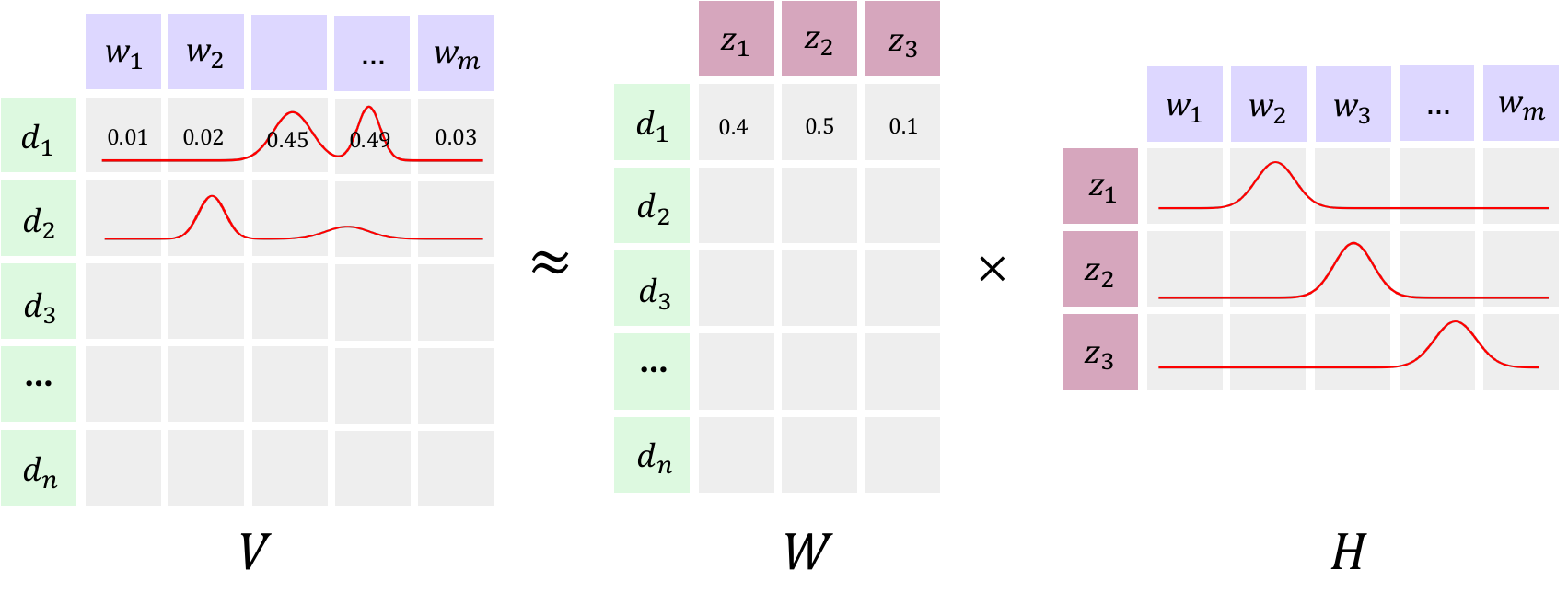}
    \vspace{-5mm}
    \caption{\revised Illustration of attention decomposition. Matrix $V$ is the attention matrix of \emph{$\text{documents } (d_i) \times \text{tokens } (w_j)$}, $W$ is the weights matrix of \emph{documents $(d_i)$ $\times$attention topics $(z_k)$}, and $H$ is each attention topic's focus of \emph{attention topics $(z_k)$ $\times$ tokens $(w_j)$}.}
    \label{fig:corpus-attn}
\end{figure}

Applying NMF to PAM's attention score across corpus involves first constructing matrix $\textbf{V}\in \mathbb{R}^{n\times m}$ to represent the attention scores of $m$ tokens across $n$ documents. Through NMF, we decompose it into matrices 
$\textbf{W}\in\mathbb{R}^{n\times k}$ and $\textbf{H}\in\mathbb{R}^{k\times m}$, where matrix $\textbf{W}$ captures $k$ topic features across $n$ documents and matrix $\textbf{H}$ contains the $k$ attention topics over $m$ tokens (Fig.\ref{fig:corpus-attn}). In our context, the matrix $\textbf{H}$ serves as the representation of corpus-level attention, where each of the $k$ topics delineates a distinct attention focus concentrated on a selective set of tokens. To optimally determine the number of topics $k$, we apply a recursive consensus clustering technique, which autonomously identifies the most suitable topic number\cite{nmf-topic-autoselect}.

\begin{figure}
    \centering
    \includegraphics[width=\linewidth]{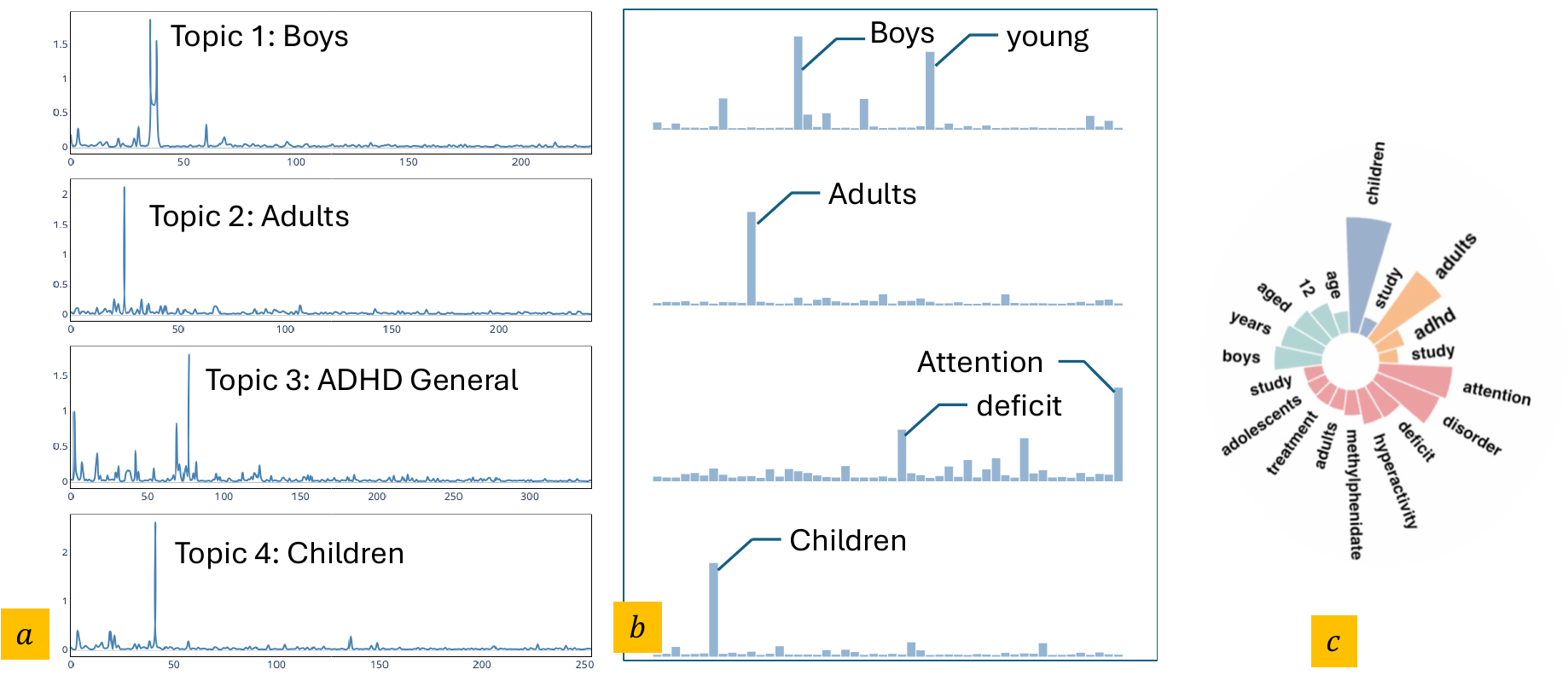}
    \vspace{-5mm}
    \caption{Illustration of attention distribution of each topic under the query ``population studied in the study'' in an ADHD corpus (2000 articles). $a$ only displays the tokens that have a decomposed value $> 0.1$, where we can clearly identify the population-related topics. $b$ is the bar chart attention visualization of the left result with the threshold $0.3$, $c$ is the circular attention wheel of the result $a$ with the threshold $0.5$.} 
    \label{fig:attn-vis}
    \vspace{-3mm}
\end{figure}

After performing the NMF for corpus-level attention, we can discern distinct topics characterized by several tokens that receive significant attention (Fig.\ref{fig:attn-vis}-a). Notably, the topics are generated based on user-provided prompts, making them more interpretable than traditional NMF outcomes that produce generic topics across the entire corpus. 
To effectively visualize this, we employ a series of bar charts where each horizontal bar represents a topic and its word distribution (Fig. \ref{fig:attn-vis}-b). This format provides clear visibility of each topic's composition and the relative attention each token receives within. We consider an alternative design is a circular bar chart that simultaneously presents multiple topics (Fig.\ref{fig:attn-vis}-c). However, this mapping make it looks like the contributions of each token across different topics is comparable, which is not necessarily the case. Additionally, using color to distinguish between topics could potentially conflict with the color coding on the document map, leading to visual confusion. Consequently, we opted for individual bar charts to maintain clarity and avoid color conflicts, ensuring a coherent visual experience.


\begin{figure}
    \vspace{-3mm}
    \centering
    \includegraphics[width=.6\linewidth]{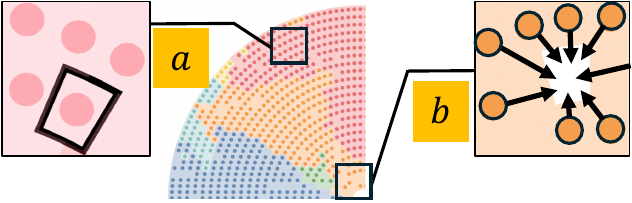}
    \caption{Coloring approach in the document map, where each cell's color is determined by its contained document's category $a$, or the most common category of its nearest 8 neighbors, if its vacant $b$.}
    \label{fig:map-coloring}
\end{figure}

\vspace{-2mm}
\subsection{Iterative Exploration in VADIS}\label{sec:iterative}
The design of our pipeline, including PAM and the subsequent document and attention visualizations, is intentionally designed to accommodate the iterative nature of information seeking. This iterative process is supported by PAM's ability to assign relevance scores and generate embeddings based on the user's specific prompt or query. In this section, we first illustrate how PAM take care of the updated prompt or new prompts, followed by the description of VADIS' interface.

\vspace{-5mm}
\subsubsection{Update Prompt or Add New Prompt} 


Considering that clinicians often need to retrieve and categorize documents based on a combined interest, we support users to generate an integrated embedding and relevance score based on multiple prompts. Specifically, when users propose multiple prompts or queries, PAM first processes each query independently, generates relevance score and the attention distribution over each document. Then PAM supports user to define a ``relevance importance weights'' to specify the important of each prompt, and calculates a composite relevance, embedding and attention distribution through a weighted sum.  

\vspace{-5mm}
\subsubsection{Visual Analytics System: \emph{VADIS}$\cdot$Insight}
\begin{figure}
    \centering
    \includegraphics[width=\linewidth]{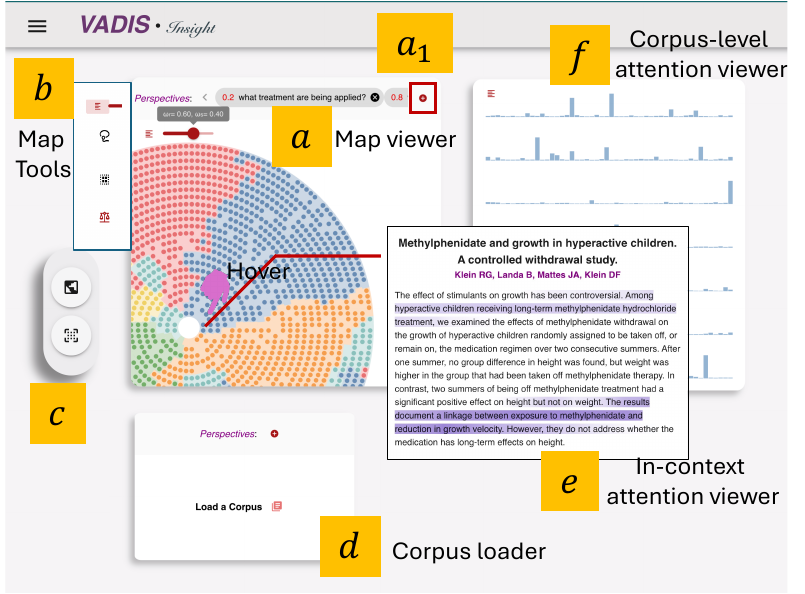}
    \caption{Interface of \emph{VADIS}$\cdot$Insight, which contains three visual components, $a$, $e$ and $f$ and support multiple interactions to facilitate iterative information-seeking.}
    \label{fig:interface}
    \vspace{-5mm}
\end{figure}
To illustrate the effectiveness and usability of our proposed pipeline, we developed a visual analytics system \emph{VADIS}$\cdot$Insight (Fig.\ref{fig:interface}), which consists of three dedicated visualization components. \emph{Firstly}, the system presents a color-coded document map based on our relevance-preserving mapping. The color of each document dot indicates the semantic similarity clusters based on query-based embeddings, and is calculated via KMeans algorithm and elbow method to automatically determine the best number of $k$.
We further color each cell in the map to make it more coherent (Fig.~\ref{fig:map-coloring}). Specifically, if a documents is assigned to a cell, the cell's color is aligned with the document's color (Fig.\ref{fig:map-coloring}-$a$), otherwise, the color of a vacant cell is determined by the most common category of its nearest 8 neighbors (Fig.\ref{fig:map-coloring}-$a$). 
\emph{Thirdly}, \emph{VADIS}$\cdot$Insight also integrates an in-context heatmap viewer to provides a detailed token-specific attention visualization within each corpus. The rationale of providing this visualization is that we consider even though the corpus-level attention can help users quickly pick up the attention overview, dive in some specific document, user can still benefit from knowing what specific token information is being attended during the 
embedding generation. 

The interface also supports multiple user interactions to facilitate the exploration. Specifically, the interface is designed as an infinite canvas. Users can seamlessly add new document maps or explore various corpus perspectives by dragging components from the VADIS menu (Fig.\ref{fig:interface}-$c$). Adding a map is straightforward—users can upload a corpus in CSV format or choose an existing one. The introduction of prompts (Fig.\ref{fig:interface}-a1) automatically generates the circular map view and triggers the attention distribution bar chart (Fig.\ref{fig:interface}-f). Interacting with the map, users can hover over documents to view in-context attention distributions. Each map is equipped with a functional menu, providing tools like lasso selection to export the selected documents into a new environment, relevance/similarity weight adjustment, and cluster quantity customization, enhancing the exploratory experience.


%% file: tex/6CaseStudy.tex
\vspace{-1mm}
\section{Evaluation}
In this section, we evaluate the performance of VADIS in information seeking. We begin by assessing PAM's effectiveness in focusing on the most relevant tokens given each query, and leveraging the attention for document retrieval and embedding generation. We then evaluate the relevance-preserving mapping, where the fitness of document-to-grid-cell relevance is quantitatively measured by the Relevance Function Correspondence metric, and validate the clustering integrity via the silhouette analysis. Finally, to illustrate the usability and effectiveness of the entire pipeline, we conduct a targeted case study with domain experts, who use VADIS to search published studies in a biomedical corpus related to ADHD research. It's also worth to note that since the corpus-level attention is designed to enhance interpretability, we mainly examine it for usability within the case study.

\begin{table}[]
\caption{Top 10 and Top 20 retrieval accuracy on test datasets, measured as the percentage of the top number of retrieved documents that contain the correct context.}
\vspace{-2mm}
\centering
\resizebox{0.99\linewidth}{!}{%
\begin{tabular}{c|cccc|cccc}
\hline
\multirow{2}{*}{\textbf{Model}} & \multicolumn{4}{c|}{\textbf{Top 10}} & \multicolumn{4}{c}{\textbf{Top 20}} \\
 &
  \multicolumn{1}{l}{emrQA} &
  \multicolumn{1}{l}{SQuAD} &
  \multicolumn{1}{l}{TriviaQA} &
  \multicolumn{1}{l|}{NaturalQA} &
  \multicolumn{1}{l}{emrQA} &
  \multicolumn{1}{l}{SQuAD} &
  \multicolumn{1}{l}{TriviaQA} &
  \multicolumn{1}{l}{NaturalQA} \\ \hline
BM25                            & 54.2    & 55.8    & 54.8    & 49.4   & 57.3    & 68.7    & 66.9   & 59.1   \\
DSSM                            & 51.9    & 52.3    & 69.4    & 70.5   & 60.9    & 63.2    & 79.5   & 78.4   \\
\multicolumn{1}{l|}{QDR}        & 68.4    & 54.1    & 76.5    & 77.1   & 73.2    & 64.5    & 82.5   & 81.0   \\
\multicolumn{1}{l|}{\textbf{\modelshort{}}} &
  \textbf{70.1} &
  \textbf{59.5} &
  \textbf{77.8} &
  \textbf{79.2} &
  \textbf{79.6} &
  \textbf{68.9} &
  \textbf{83.3} &
  \textbf{84.9} \\ \hline
\end{tabular}%
}
\label{tab:total-attn evaluation}
\vspace{-2mm}
\end{table}

{ \revised
\subsection{Evaluation of Prompt-based Attention Model}
PAM is trained on question-answering datasets to assign higher attention scores to tokens relevant to the prompt. The evaluation focuses on the correlation between prompts and generated embeddings at document and token levels. At the document level, the correlation is first measured by evaluating PAM’s ability to assign higher relevance scores to contexts containing the answer. We then assess the quality of the generated embeddings by showcasing their ability to cluster under different user interests. At the token level, PAM ensures higher attention scores for important tokens that form the answer to the query.



\vspace{-1mm}
\subsubsection{Document-Level Attention Evaluation}We evaluate this by computing the relevance scores of all documents for each query. We then extract the top-10 and top-20 contexts ranked by these relevance scores and compute the accuracy at which the ground-truth context is included. 
We chose BM25\cite{bm25}, DSSM\cite{DSSM} and QDR\cite{Docflow} as baseline models for comparison. 
The results are reported in the \autoref{tab:total-attn evaluation}, where we observe that PAM achieves higher accuracy in all four test datasets, especially on emrQA. We believe \modelshort{}'s superior performance on the emrQA dataset is because we use BioBERT to initialize both the $E_P(\cdot)$ and $E_D(\cdot)$. 
In addition, PAM performs better than QDR, which also uses dual-encoders for document retrieval. We infer that the reason is PAM leverages the embeddings of all tokens in each document to calculate the relevance score, while QDR only uses the embedding of [CLS] token to represent the context.
\vspace{-1mm}
\subsubsection{Document Embedding Correlation Evaluation}
To further assess the correlation between prompts and generated embeddings, we construct a dataset of 2,000 literature related to ADHD studies, each labeled according to the population being studied (e.g. children, teenagers, adults), treatment applied and other treatment for comparison. We calculate the embeddings for each document given these three corresponding prompt and apply k-means clustering, with $k$ equal to the number of groups labeled for each prompt. The clustering accuracy is then evaluated using the Adjusted Rand Index (ARI) \cite{ARIs}, which quantifies the agreement between the clustering result and the true labels, and ranges from -1 to 1, with scores close to 1 indicating excellent clustering performance, 0 being random and -1 being completely different. For comparison, we report the performance of BioBERT embeddings without PAM. 
As shown in Table \ref{tab:ari}, PAM achieves significantly better clustering performance, highlighting its effectiveness in capturing prompt-specific relevance compared to static BioBERT embeddings. We also include the visualization result of the clustering in the supplemental material.
\begin{table}[]
\centering
\caption{\revised Clustering accuracy evaluated using adjusted rand index (ARI) for different prompts.}
\vspace{-2mm}
\resizebox{0.7\linewidth}{!}{%
\begin{tabular}{c|c|c}
\hline
\textbf{Prompt} & BioBERT & PAM \\ \hline
Population being studied (5 groups)            & 0.26   &\textbf{0.71}                               \\ \hline
Treatment being applied (7 groups)           & 0.41    &\textbf{0.79}                         \\ \hline
Treatment being compared (6 groups)         & 0.14     &\textbf{0.62}                          \\ \hline
\end{tabular}%
}
\vspace{2mm}
\label{tab:ari}
\vspace{-5mm}
\end{table}

\vspace{-1mm}
\subsubsection{Token-level Correlation Evaluation}
To evaluate the token-level attention, we feed each $<$query, context$>$ into PAM and obtain the attention scores across all tokens in the context. 
To quantify the proportion of attention assigned to the relevant answer tokens, we define a metric called \textit{Relevance Attention Ratio (RAR)}. It is computed as the ratio of attention scores on answer tokens to the total attention scores across all tokens. A higher RAR indicates the PAM attends more to the most relevant information within the context. The RAR results on four test datasets are reported in \autoref{tab:rar}. 
sOn average, \modelshort{} allocates about 55.8\% of its attention to answer tokens across the four test datasets. This suggests PAM  captures both the answer information itself and the surrounding context, while focusing more on the directly relevant answer tokens. It is important to note that the \modelshort{} is not directly supervised to only attend to answer tokens, but is trained to assign proper attention for retrieving relevant documents. The results indicate this training strategy works well for the task. 
}

\begin{table}[]
\centering
\caption{Averaged relevance attention ratio (measured in percentage) of \modelshort{} on four test datasets.}
\vspace{-2mm}
\resizebox{0.5\linewidth}{!}{%
\begin{tabular}{c|c}
\hline
\textbf{Dataset} & \textbf{Relevance Attention Ratio} \\ \hline
emrQA            & 62.1                               \\ \hline
SQuAD            & 53.8                               \\ \hline
TriviaQA         & 46.9                               \\ \hline
NaturalQA        & 60.6                               \\ \hline
\end{tabular}%
}
\vspace{2mm}
\label{tab:rar}
\vspace{-8mm}
\end{table}

\begin{figure*}
    \centering
    \includegraphics[width=\linewidth]{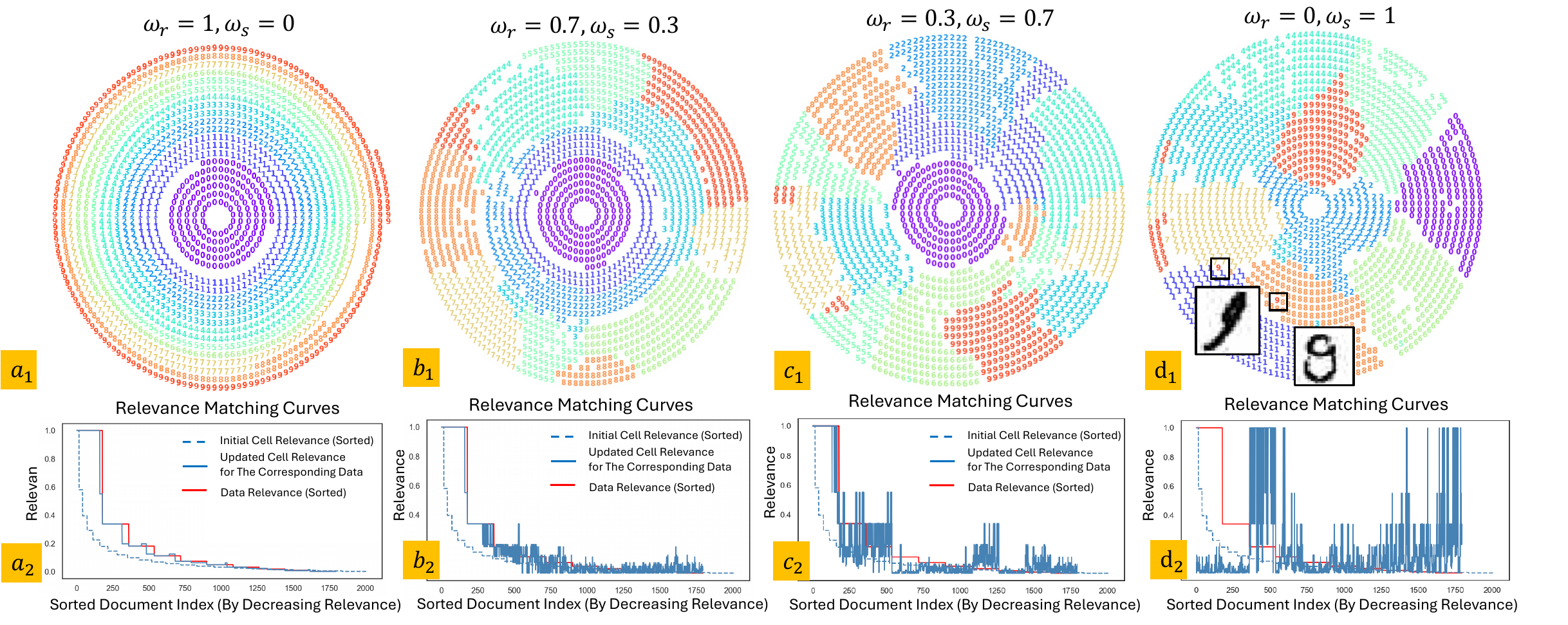}
    \vspace{-5mm}
    \caption{The mapping of MNIST dataset of 1791 samples with various relevance $\omega_r$ and similarity $\omega_s$ pairs. $a_1$, $b_1$, $c_1$ and $d_1$ are the projected map for various relevance and similarity weights; $a2$, $b2$, $c2$ and $d2$ are the relevance profile of the data and the change of corresponding cells' relevance before and after the training.} 
    \label{fig:map-evaluate}
    \vspace{-2mm}
\end{figure*}

\vspace{-1mm}
\subsection{Evaluation of Relevance-preserving Mapping}
The relevance-preserving mapping is proposed to project documents based on two critical dimensions: relevance and similarity. We also highlight its inherent flexibility for utilization across various data types. To demonstrate this versatility and ensure clarity in our evaluation, we select the MNIST dataset to facilitate an intuitive understanding of the algorithm's effectiveness. Specifically, We assign relevance scores inversely proportional to digit values as $\frac{1}{digit + 1}$, with lower digits indicating higher relevance, and use a pixel-based feature for similarity measurement. This setup makes the projection result easy and intuitive to interpret, both from the relevance and similarity perspectives.

To quantitatively measure the algorithm's performance, we introduce Relevance Profile Correspondence (RPC) to assess relevance alignment and the silhouette coefficient for clustering quality. 
RPC is calculated by measuring the distance between two key profiles: the inherent relevance distribution of the dataset and the attributed relevance distribution of the grid cells post-training. As illustrated in Fig.\ref{fig:map-evaluate}-$a_1$, initially, each cell's relevance is set based on its proximity to the center, assuming a radial gradient of relevance, and through the training process, these cell relevance are dynamically adjusted to more closely mirror the actual dataset's relevance distribution. This quantitative measure offers a precise evaluation of the mapping algorithm's capacity to preserve the relevance hierarchy within the visual layout.

\begin{figure}
    \vspace{-4mm}
    \centering
    \includegraphics[width=0.9\linewidth]{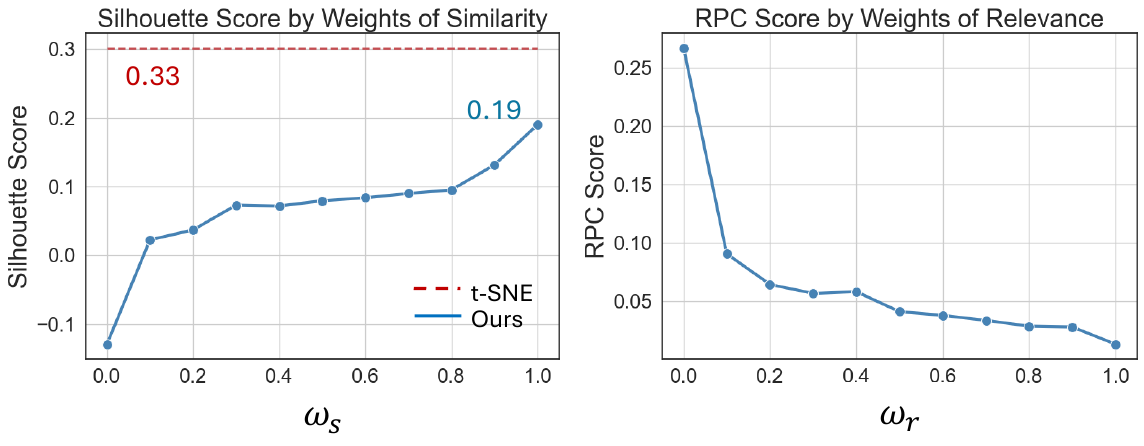}
    \vspace{-3mm}
    \caption{Evaluation of relevance-preserving mapping on 1797 MNIST samples using silhouette scores over similarity weights $\omega_s$ and RPC scores over relevance weights $\omega_r$. }
    \label{fig:res-metrics}
    \vspace{-7mm}
\end{figure}

We adopt a variety of configurations of relevance ($\omega_r$) and similarity ($\omega_s$) weights to assess the mapping algorithm's precision in integrating dual criteria. The silhouette analysis (Fig.\ref{fig:res-metrics}) measures the cohesion and separation of clusters and with higher values indicate better-defined clusters. The result in Fig.\ref{fig:res-metrics} demonstrates a peak score of 0.2 under maximum similarity weighting, juxtaposed with t-SNE’s silhouette score of 0.33. Despite the difference, we consider that our approach is distinct in its capacity to integrate relevance into the clustering process, thus serving divergent analytical objectives.
The relevance of this distinction is further exemplified in Figure \ref{fig:map-evaluate}, which presents two contrasting scenarios ($a1$ and $d1$) through the lens of RPC. When relevance is exclusively considered ($\omega_r=1,\omega_s=0 $), the post-training cell relevance closely mirrors the actual data relevance and arranging digits sequentially from the center outwards, indicating a precise alignment. Inversely, when similarity is solely prioritized ($\omega_r=0,\omega_s=1  $), there is a discernible divergence between cell and data relevance, however the clusters in this case corresponds closely to distinct digits, but with anomalies like a '9' within the '8' cluster emerge. Our detailed inspection reveals its visual similarity to '8'. The intermediate cases, with relevance and similarity weights set to ($\omega_r=0.7, \omega_s=0.3$) and ($\omega_r=0.3, \omega_s=0.7$) respectively (Fig.\ref{fig:map-evaluate}-$b_1,b_2,c_1,c_2$) shows a balance between two criteria. With a greater weight on relevance ($\omega_r=0.7$), the map still maintains a clear gradient of relevance from center outwards while beginning to form semantic clusters. Conversely, when similarity is weighted more heavily ($w_s=0.7$), the semantic clusters become more pronounced, yet the most relevant digits (digit `0`) are still centrally located.

These findings collectively substantiate the versatility of our relevance-preserving mapping algorithm, demonstrating its robust performance across various configurations and its aptitude for generating insightful visualizations tailored to specific user-defined parameters.
 
\vspace{-2mm}
\subsection{Case Study} 
This study aimed to demonstrate the usability of VADIS in an efficient corpus analysis and interactive information-seeking for relevant documents. Domain expert E1 was invited to participate in the case study. Motivated by a recently published systematic review \cite{nanda2023adverse} on the side effects of methylphenidate, E1 sought to broaden the investigation to other ADHD treatments and their side effects. 

To achieve this, E1 first prepared a dataset of $2108$ article abstracts by searching
the terms ``ADHD, treatment and side effects across different populations.''
E1 proceeded to upload the dataset to VADIS.
With an initial interest in different study population groups in the corpus. E1 entered the prompt ``populations being studied.''
Understanding that all documents are likely to reference some population, 
she set the similarity weights $\omega_s$ to 1 and relevance weights $\omega_r$ to 0, aiming to emphasize similarity in the analysis. 
VADIS processed the prompt and generated a document map with six distinct clusters (Fig.\ref{fig:teaser}-$b$). 
This visualization allowed E1 to interactively explore each cluster by hovering over individual articles.  
Such exploration provided insights into the details of each article, including the tokens that garnered the most attention during the embedding process. 
This examination revealed clusters specifically discussing adults (\tikzcircle[fill={rgb,255:red,126; green,176; blue,126}]{3pt}), children (\tikzcircle[fill={rgb,255:red,171; green,209; blue,206}]{3pt}, \tikzcircle[fill={rgb,255:red,243; green,140; blue,146}]{3pt}), healthy subjects and volunteers (\tikzcircle[fill={rgb,255:red,128; green,196; blue,187}]{3pt}), general patient information (\tikzcircle[fill={rgb,255:red,248; green,158; blue,87}]{3pt}), and boys varied age groups (\tikzcircle[fill={rgb,255:red,250; green,225; blue,168}]{3pt}). After confirming these study populations, E1 was interested in understanding the model's global attention focus in the corpus. By engaging with the corpus-level attention distribution (Fig.\ref{fig:teaser}-$c$), six topics emerged. Among these, E1 was able to find that there were two topics addressing general ADHD information, while the other four focused on specific population segments previously identified. 

Building upon the insights gained from the corpus-level visualization of population groups, E1 shifted her focus toward understanding the diversity of treatments discussed within the corpus. To this end, she refined the query to investigate ``treatments being studied,'' maintaining the semantic weights at $1$ and relevance weights at $0$ to prioritize semantic similarity in the document distribution. Upon updating the query, VADIS updated the document map and displayed seven distinct clusters (Fig.\ref{fig:teaser}-$d$). This nuanced visualization enabled the domain expert to identify specific clusters dedicated to various treatments: one cluster detailed behavioral therapy (\tikzcircle[fill={rgb,255:red,248; green,158; blue,87}]{3pt}), another discussed methylphenidate (\tikzcircle[fill={rgb,255:red,126; green,176; blue,126}]{3pt}), a third was centered on placebo use (\tikzcircle[fill={rgb,255:red,128; green,196; blue,187}]{3pt}), another focused on stimulants (\tikzcircle[fill={rgb,255:red,250; green,225; blue,168}]{3pt}), and yet another on atomoxetine (\tikzcircle[fill={rgb,255:red,243; green,140; blue,146}]{3pt}). This clear clustering among treatments offered a comprehensive view of the treatments studied within the corpus. Further exploration through the corpus-level attention visualization confirmed the presence of two broad topics related to the general ADHD information and five dedicated topics that elucidated different treatments ((Fig.\ref{fig:teaser}-$e$)). However, E1 noted that the current distribution of attention did not encompass details on the side effects or the effectiveness of these treatments. E1 then prepared to delve deeper into the side effects associated with the identified treatments, marking a transition to the next phase of her systematic review process. 

Following the exploration of treatment types, the domain expert proceeded to input a secondary query concerning ``side effects of the treatments,'' adjusting the weight of treatments to 0.8. This modification aimed to balance the focus between treatment types and their side effects within the document representation. VADIS generated an updated document map, showcasing a refined distribution of articles (Fig.\ref{fig:teaser}-$a$). This updated map revealed a distinction between articles: several were positioned closer to the center, indicating a higher relevance to the query, especially regarding side effects, while others were located more peripherally. Through interactive examination of the articles, the domain expert observed that clusters were organized around specific treatments. Notably, articles nearer the center predominantly discussed side effects, providing a clear demarcation of their importance in the analysis. For instance, articles about methylphenidate prominently mentioned side effects such as headaches and loss of appetite, while those about atomoxetine frequently reported headaches. This organized visualization facilitated more structured navigation through the corpus, allowing the domain expert to systematically review articles from those centrally located, which detailed side effects, outward to those focusing on treatments in general. In the concluding phase of the case study, the domain expert expressed interest in locating the studies cited in the original systematic review on the document map. Upon marking these studies (Fig.\ref{fig:teaser}-$a \star$), it was observed that they were closely situated near the center of the map. This demonstrated the VADIS' effectiveness in identifying articles of critical relevance, particularly those identified in the systematic review, affirming the robustness of VADIS in facilitating targeted and insightful exploration of biomedical literature.


%% file: tex/7_discussion.tex
\vspace{-2mm}
\section{Discussion and Limitation}
{\revised
\noindent\textbf{Fast inference of \frameworkshort{}.} The design of the PAM defers the involvement of the prompt embedding until the final attention layer. This allows the document encoder to preprocess the document collection and store the token embeddings of each document before the start of an information-seeking process. By doing so, the prompt-based document representation only needs to go through the prompt-encoding and final attention layer during inference, which can significantly reduce the time needed for real-time inference. 
}

\noindent\textbf{Enhancing \modelshort{}'s performance with diverse QA datasets.} While our primary focus is on the biomedical domain, and emrQA is most relevant, incorporating general datasets such as SQuAD, TriviaQA, and Natural Questions is still beneficial since many clinical queries are expressed in general question formats, like ``what's the population being studied," which can be learned from general QA datasets. 
We provide a detailed performance analysis with the involvement of different training datasets in the supplemental material.

\noindent\textbf{Transferability and extensibility of VADIS.}
In this paper, we mainly demonstrate the usage of VADIS in the biomedical domain. However, it can also benefit other domains by replacing the $E_D(\cdot)$ with other domain-specific language model. 

{\revised
\noindent\textbf{Limitation. }The relevance-preserving map encounters a limitation due to the intrinsic trade-off between relevance and similarity. This balance can sometimes results in irregular empty spaces and sub-optimal document alignment. We consider this reflect the intrinsic challenges of the SOM's competitive learning mechanism. While SOM is adept at projecting data from continuous high dimension to discrete grid, it cannot guarantee an optimal placement for all documents, therefore limit the map from achieving an optimal alignment between similarity and relevance across the corpus. Additionally, PAM faces limitations when prompt is less relevant to the corpus; in such cases, the attention score assigned to the tokens are generally small and lack significant focus. When these small attention scores are softmaxed to calculate embeddings, they lead to embeddings that broadly cover all semantics of the token information rather than highlighting specific relevance.
}
\vspace{-3mm}
\section{Future work and conclusion}
\noindent\textbf{Future work.}
In future work, we aim to enhance the VADIS framework to automate the selection of relevance and similarity weights, which currently depend on manual input. By incorporating algorithms that adjust these weights based on the data relevance, the relevance-preserving map will offer optimized document map layouts, improving user experience and efficiency. 

\noindent\textbf{Conclusion.}
In this paper, we propose a visual analytics pipeline for dynamic document representation and information seeking. Our approach incorporates the Prompt-based Attention Model to generate user-driven document embeddings and facilitate efficient exploration of a corpus with a novel relevance-preserving document map. Our pipeline further integrates a corpus-level attention visualization to enhance the interpretability of the model's focus. This integrated solution empowers researchers to
better navigate and understand large document collections, ultimately enhancing the overall research process.

